\documentclass[aps,pra,twocolumn,amsmath,amssymb,superscriptaddress,showpacs]{revtex4}
\usepackage{longtable}
\usepackage{graphicx}
\usepackage{amsmath}
\usepackage{multirow}
\usepackage{layout}

\newcommand{\ImSize} {8.5cm}     

\begin{document}
\author{I. A. Sulai}
\affiliation{Physics Division, Argonne National Laboratory, Argonne, Illinois 60439, USA}
\affiliation{Department of Physics,The University of Chicago, Chicago, Illinois 60637, USA}
\affiliation{The Enrico Fermi Institute, The University of Chicago, Chicago, Illinois 60637, USA}
\author{Qixue Wu}
\affiliation{Department of Physics, University of Windsor, Windsor, ON N9B 3P4, Canada}
\author{M. Bishof}
\affiliation{Physics Division, Argonne National Laboratory, Argonne, Illinois 60439, USA}
\affiliation{The Enrico Fermi Institute, The University of Chicago, Chicago, Illinois 60637, USA}
\author{G. W. F. Drake}
\affiliation{Department of Physics, University of Windsor, Windsor, ON N9B 3P4, Canada}
\author{Z.-T. Lu}
\affiliation{Physics Division, Argonne National Laboratory, Argonne, Illinois 60439, USA}
\affiliation{Department of Physics,The University of Chicago, Chicago, Illinois 60637, USA}
\affiliation{The Enrico Fermi Institute, The University of Chicago, Chicago, Illinois 60637, USA}
\author{P. Mueller}
\affiliation{Physics Division, Argonne National Laboratory, Argonne, Illinois 60439, USA}
\author{R. Santra}
\affiliation{Department of Physics,The University of Chicago, Chicago, Illinois 60637, USA}
\affiliation{Chemical Sciences and Engineering Division, Argonne National Laboratory, Argonne, Illinois 60439, USA}

\date{\today}
\title{Hyperfine Suppression of $2~^3{\rm S}_1 ~-~ 3~^3{\rm P}_J$ Transitions in $^3$He}
\begin{abstract}

Two anomalously weak transitions within the $2\;^3{\rm S}_1~-~3\;^3{\rm P}_J$
manifolds in $^3$He have been identified. Their transition strengths
are measured to be 1,000 times weaker than that of the strongest transition
in the same group. This dramatic suppression of transition strengths is due
to the dominance of the hyperfine interaction over the fine structure
interaction. An alternative selection rule based on \textit{IS}-coupling (where the nuclear spin is first coupled to the total electron spin) is proposed.
This provides qualitative understanding of the transition strengths.
It is shown that the small deviations from the \textit{IS}-coupling model are fully
accounted for by an exact diagonalization of the strongly interacting
states.
\end{abstract}

\maketitle

Persistent efforts in both theory and experiment have yielded
increasingly precise understanding of the helium atom. Due to its
simplicity, the helium atom has been a proving ground for precision
atomic measurements and calculations of few-body quantum systems.
The knowledge gained from this effort is used to test bound-state
quantum electrodynamics \cite{Morton1,Morton2,Pachucki}, determine
the fine structure constant \cite{Zelevinsky,George}, and explore exotic
nuclear structure \cite{WangPRL,Mueller,Shiner,Marin}. We report
results of a combined theoretical and experimental study on the
strengths of $2\;^3{\rm S}_1 - 3\;^3{\rm P}_J$ transitions in $^3$He.

\begin{figure}[h!]
\includegraphics[width=\ImSize]{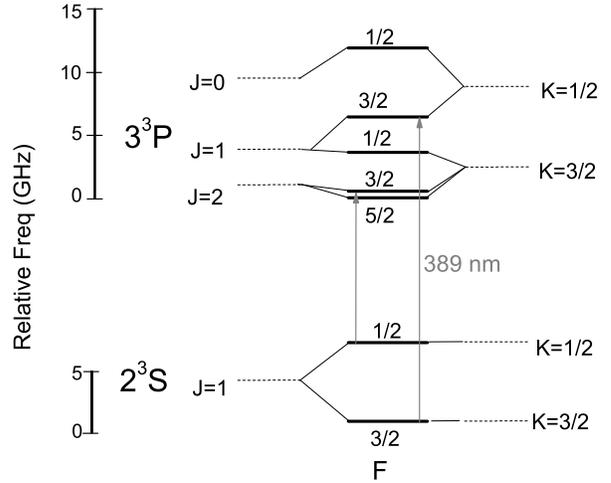}
\caption{Level scheme of  $^3$He showing the levels investigated,
with the arrows indicating the suppressed transitions observed. The
level positions are drawn to scale within each manifold. The large
hyperfine splitting with respect to the fine structure splitting is
evident. With a nuclear spin $I = 1/2$ for $^3$He, the
levels are designated by the familiar term symbols, with
$\mathbf{J} = \mathbf{L}+\mathbf{S}$, $\mathbf{F} = \mathbf{J} +
\mathbf{I}$ on the left. The levels are labeled on the right using
the quantum number $\mathbf{K} = \mathbf{I} + \mathbf{S}$,
$\mathbf{F} = \mathbf{K}+\mathbf{L}$.}
\label{fig1}
\end{figure}

Surprisingly, we observe that the strengths of two ``allowed''
transitions, $2\;^{3}{\rm S}_{1},(F=\frac{3}{2})~-~3\;^{3}{\rm P}_{1},(F=\frac{3}{2}) $ and
$2\;^3{\rm S}_1,(F = \frac{1}{2})~-~3\;^3{\rm P}_2,(F = \frac{3}{2})$, are 1,000 times weaker than that of the
strongest transition $2\;^3{\rm S}_1, (F = \frac{3}{2})~-~3\;^3{\rm P}_2,(F = \frac{5}{2})$. The level scheme showing these transitions is presented in Fig \ref{fig1}. This
dramatic suppression of transition strengths is due to a rare
atomic phenomenon: within the $3\;^3{\rm P}$ manifold, the hyperfine
interaction is comparable to or even stronger than the fine
structure interaction. Consequently the conventional model based on
\textit{LS}-coupling is no longer applicable. Rather, we find that an
alternative model where the fine structure interaction is treated as a
perturbation on states obtained by first coupling nuclear spin to the total electron
spin provides a good qualitative explanation of the observed
suppression. We refer to this coupling scheme as \textit{IS}-coupling. We
start by discussing the details of the experiment and compare the
data with the predictions from the different coupling schemes.
Finally, we discuss an exact diagonalization method to account for
the small differences between experiment and the \textit{IS}-coupling
scheme.

We measure the ratio of transition strengths using a cross-beam
laser induced fluorescence method. A beam of metastable helium atoms
in the $2\;^3{\rm S}_1$ state is prepared in a liquid-nitrogen cooled
RF-driven discharge. A retro-reflected beam of linearly polarized
389 nm light is incident perpendicular to the atomic beam. The
polarization of the light is along the direction of the atomic beam.
A uniform external magnetic field of 5 Gauss is applied along the
direction of the laser to provide an axis of quantization. As the
frequency is scanned across different resonances, the atoms are
excited, and fluorescence from the atoms is detected in the
direction normal to the atomic and laser beams. The metastable
atomic beam is collimated using a collimator, made of a stack of
microscope cover slips which provides high collimation in the
direction along the laser beam  \cite{Tanner}. We are able to obtain
Doppler broadened lines of $20$ MHz linewidth. The natural linewidth
of the transitions is 1.6 MHz. Approximately 4 mW of 389 nm light is
obtained by frequency doubling infrared light at 778 nm. The
frequency of the 778 nm light is referenced to a temperature
stabilized Fabry-Perot cavity. The power of the laser and its
wavelength are monitored continuously.

\begin{figure}[h!]
\includegraphics[width=\ImSize]{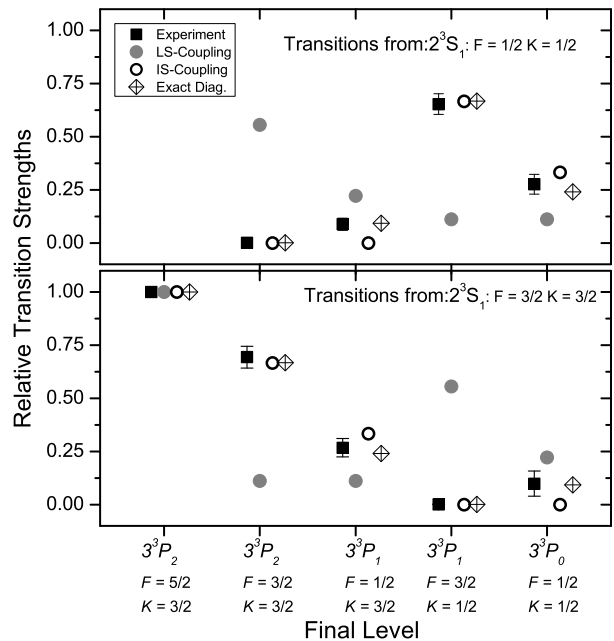}
\caption{Comparison of relative transition strengths for all E1 allowed
transitions between the $2\;^3{\rm S}_1$ and $3\;^3{\rm P}_J$ manifolds. All
values are normalized with respect to the $2\;^3{\rm S}_1, (F=\frac{3}{2}) - 3\;
^3{\rm P}_2, (F=\frac{5}{2})$ transition.} \label{fig2}
\end{figure}

The nine E1 allowed transitions are repeatedly probed in a random
order and the spectra are recorded. Each spectrum is fitted using a
statistically weighted Voigt profile. The integrated area of the
profile divided by the power of the probing laser beams is taken as
a measure of the transition strength. As the absolute
atomic beam flux and efficiency of detecting the fluorescence
photons are not measured in this experiment, only the ratios of transition strengths are determined. By
defining the strength of the strongest transition, $2\;^3{\rm S}_1, (F = \frac{3}{2})~-~3\;^3{\rm P}_2,(F = \frac{5}{2})$, to be unity, we determine the relative
strengths of the other eight transitions. The results are presented
in Fig. \ref{fig2} and in Table \ref{Data}.

The intensity of the probing laser beam is varied depending on the
transition under study. For example, when probing the two highly
suppressed transitions, the intensity of the probe is increased by
two orders of magnitude. In all cases, however, the laser intensity
is kept well below the saturation intensity of the particular
transition under study. Indeed, the intensity is chosen so that on
average less than one photon is scattered by each atom as it passes
the laser beams in approximately $2~ \mu$s. This is to avoid
nonlinear effects in the measurements due to optical pumping and
mechanical effects of the light on the atomic beam. Such systematic
effects are studied by examining the dependence of transition signal
on the laser beam power. Additional corrections are made and
systematic errors generated due to changing background in the
measured laser power and the anisotropic angular distribution of
the fluorescence emission. The final error estimates are given in
Table \ref{Data}.

\begingroup
\begin{table}[ht!]
\caption{Relative transition strengths for all E1 allowed
transitions between the $2\;^3{\rm S}_1$ and $3\;^3{\rm P}_J$ manifolds. All
values are normalized with respect to the $2\;^3{\rm S}_1, (F=\frac{3}{2}) - 3\;
^3{\rm P}_2, (F=\frac{5}{2})$ transition.}
\begin{ruledtabular}
\begin{tabular}{ccllll}
Initial (J,F)   & Final (J,F)&  Experiment & LS & IS  & Exact     \\
 $2\;^3{\rm S}_J$& $3\;^3{\rm P}_J$ & & & & Diag.\\
\hline
\multirow{5}{*}{(1,3/2)} & (2,5/2)&   1  & 1 & 1 & 1\\
   & (2,3/2)&   0.69(5) & 0.11 & 0.67 & 0.67\\
   & (1,1/2)&   0.26(4) & 0.11 & 0.33 & 0.24\\
   & (1,3/2)&   0.0012(2) & 0.55 & 0 & 0.0010\\
   & (0,1/2)&   0.10(5) & 0.22 & 0 & 0.093\\
\hline
 \multirow{4}{*}{(1,1/2)}& (2,3/2)&   0.0011(4) & 0.55 & 0 & 0.0010\\
  & (1,1/2)&   0.08(3) & 0.22 & 0 & 0.093\\
  & (1,3/2)&   0.65(4) & 0.11 & 0.67 & 0.67\\
   & (0,1/2)&   0.27(4) & 0.11 & 0.33 & 0.24\\
\end{tabular}
\end{ruledtabular}
\label{Data}
\end{table}
\endgroup

The textbook strategy \cite{Sobelman} to estimate theoretically the 
atomic transition strengths is based on the presumed hierarchy that hyperfine splittings be small in
comparison with fine-structure splittings.  Consequently approximate
eigenstates of the total Hamiltonian may be constructed by first
coupling $L$ (total orbital angular momentum quantum number) and $S$
(total electronic spin quantum number) to form the total electronic
angular momentum $J$; coupling $J$ and $I$ (nuclear spin quantum
number) then gives the total atomic angular momentum $F$. Within
this \textit{LS}-coupling model, the total strength for an electric dipole
transition may be evaluated using standard angular momentum algebra
\cite{Edmonds}.

The results of this $LS$-coupling model are compared with the
experimental data in Fig.~\ref{fig2}.  It is apparent that there is
not even qualitative agreement. The origin of the failure of the
$LS$-coupling model may be understood as follows. In $^3$He, the
hyperfine structure is almost entirely due to the magnetic dipole
interaction of the tightly bound $1s$ electron with the nucleus. The
fine structure is a consequence of both one-body spin-orbit coupling
of the excited $nL$ electron and two-body spin-other-orbit and
spin-spin interactions of the $nL$ electron with the $1s$ electron
\cite{BetheSalpeter}.  As $n$ increases, the fine-structure
splittings decrease as $n^{-3}$.  The hyperfine interaction of the
$1s$ electron, on the other hand, tends for large $n$ to the constant
hyperfine interaction strength in $^3$He$^+$. Note that the
hyperfine splitting in the ground state of $^3$He$^+$ is $8.7$~GHz
\cite{FoMa66}, which is comparable to, or larger than, the level
spacings within the $2$~$^3S$ and $3$~$^3P$ manifolds (see
Fig.~\ref{fig1}).

The relative strength of the hyperfine interaction in $^3$He has
been recognized before
\cite{FrTo51,TiAn73,DeLo80,FrLi80,BlCo82,PrHi83,Vassen} and has been
taken as an indication that a simple angular momentum coupling model
describing transitions in $^3$He is not available and that a
numerical diagonalization of an effective Hamiltonian is necessary.
We demonstrate in the following that although $n$ is quite small in
the ${\rm 3}$~$^3{\rm P}$ manifold, the assumption of relatively weak fine
structure interactions does provide a simple model that allows us to
understand qualitatively the strengths of transitions from $2$~$^3{\rm S}$
to $3$~$^3{\rm P}$.

For $3\;^3{\rm P}$, ${\rm S}$ is still a good quantum number, since the
separation of this manifold from $3\;^1 {\rm P}$ is large ($\sim
10^4$~GHz) in comparison with the hyperfine and fine structure
splittings.  Therefore, the basic idea underlying what we refer to
as the $IS$-coupling model is that the electrostatic exchange
interaction between the two electrons preserves $\bf{S}$; the
hyperfine interaction couples $\bf{S}$ and $\bf{I}$ to form a new intermediate
angular momentum $\bf{K}$; and $\bf{F}$ is then obtained by coupling $\bf{L}$ and $\bf{K}$.
In this picture, the $^3$He eigenstates of relevance here are not
labeled in terms of $nLS(J)I,F$, but in terms of $nIS(K)L,F$. An
immediate consequence of the fact that the electric dipole operator
acts on neither $S$ nor $I$ is that $K$ must be conserved in an $E1$
transition, i.e.,
$|\langle\Psi^{(n^{\prime}L^{\prime}S^{\prime}I)}_{K^{\prime}F^{\prime}}
\parallel \hat{D}\parallel\Psi^{(nLSI)}_{K F}\rangle|^2$ vanishes if
$K$ differs from $K^{\prime}$. A similar model was used in 1933 for
a case in which $S$ is not conserved \cite{GoBa33}, but that appears
to be the only other study employing an extreme hyperfine-coupling
picture to develop a basic understanding of transition strengths
involving hyperfine multiplets.

As shown in Fig.~\ref{fig2}, there is good qualitative agreement
between experiment and the $IS$-coupling model, thus suggesting that
already for $n=3$, the fine-structure interactions may be considered
perturbations to the hyperfine structure.  For instance, within the
$IS$-coupling model, the suppression of the transition from
$2$~$^3{\rm S}_1$, $F=\frac{3}{2}$ ($K=\frac{3}{2}$) to $3$~$^3{\rm P}_1$, $F=\frac{3}{2}$ ($K=\frac{1}{2}$)
follows from the $K$-selection rule in $E1$ transitions. On the
other hand, according to experiment, the transition from
$2$~$^3{\rm S}_1$, $F=\frac{3}{2}$ ($K=\frac{3}{2}$) to $3$~$^3{\rm P}_0$, $F=\frac{1}{2}$ ($K=\frac{1}{2}$) is
weakly allowed, in slight deviation from the $IS$-coupling model.
We note that the observed suppressions for certain transitions from
$2$~$^3{\rm S}$ to $5$~$^3{\rm P}$ \cite{BlCo82,Vassen} are fully consistent
with the $K$-selection rule.

In order to characterize the nature of the perturbations to the
$IS$-coupling model for $^3$He, and account for the slight deviations,
we have performed an exact
diagonalization of the total Hamiltonian $H$ within the manifold of
$3\;^3{\rm P}$ and $3\;^1{\rm P}$ states, including both fine and
hyperfine structure.  The total Hamiltonian of $^{3}$He in the
absence of external fields is of the form
\begin{eqnarray}
H= H_{\rm NR} + H_{\rm{fs}}+H_{\rm{hfs}}
\end{eqnarray}
where $H_{\rm NR}$ is the nonrelativistic Hamiltonian, $H_{\rm{fs}}$
represents the fine structure interaction for helium as described by
many authors (see Drake \cite{a1,a2} for a review), and $H_{\rm
hfs}$ represents the hyperfine structure interaction, see for
example, Bethe and Salpeter \cite{BetheSalpeter}. In this picture, $H_{\rm
hfs}$ is treated as a small perturbation relative to the large
electrostatic splitting between states with different principal
quantum number $n$, and by exact diagonalization within the manifold
of strongly mixed states with the same $n$.  The technique is
basically the same as that described by Hinds, Prestage and
Pichanick \cite{a5}.

Using these methods, a comprehensive investigation of the fine and
hyperfine structure of $^{3}$He has recently been carried out by
Morton, Wu, and Drake \cite{Morton2}.  All  fine structure and hyperfine
structure parameters required to diagonalize the complete fine and
hyperfine interaction matrix in the basis set of singlet and triplet
states are accurately calculated by using  double basis set
variational wave functions in Hylleraas coordinates as described by Drake
\cite{a1,a2}. For the $3P$ state, instead of using directly the
theoretical energies for $^3$He, we combined the
theoretical isotope shifts for $^{3}$He relative to $^{4}$He \cite{Morton2}
with the best experimental ionization energies for $^{4}$He.\cite{Mueller,Morton2}.
This gives higher accuracy  due to cancelations of the
mass-independent QED uncertainties in the calculated isotope shifts.

The final step is to calculate the electric dipole transition line
strengths between
the perturbed  hyperfine states of $2\;^3{\rm S}$ and $3\;^3P$ in terms of
standard angular momentum theory, in which the perturbed hyperfine
states are linearly expanded in terms of unperturbed eigenstates.
The expansion coefficients are obtained by the above diagonalization
of the complete matrix. The final results and the comparison with
experiment are given in Table \ref{Data}. The
calculations show that the  mixing between hyperfine states of $3\;^3{\rm P}$
with different $K$ but the same $F$ of  $^{3}$He precisely accounts for
the deviations shown in Table \ref{Data} from the \textit{IS}-coupling model. This mixing is due to the fine structure interactions. We find that both the one-body spin-orbit, and the two-body spin-spin and spin-other-orbit interactions must be included, in order to accurately reproduce the strengths. 
In the case of the 2$S$ state, this hyperfine mixing is also important for
hyperfine structure, as shown by Riis \textit{et al.} \cite{a7}, but its
contribution to the transition strength is negligible in the present
work.

In summary, the hyperfine suppression of $2\;^{3}{\rm S}_{1},(F=\frac{3}{2}) $ to
$3\;^{3}{\rm P}_{1},(F=\frac{3}{2})$ and $2\;^{3}{\rm S}_{1},(F=\frac{1}{2}) $ to
$3\;^{3}{\rm P}_{2},(F=\frac{3}{2})$ radiative transitions in $^3$He
is caused by a selection rule that emerges in the limit of strong hyperfine
mixing between states with the same $F$ but different $J$.  In this
limit, the radiative transitions are better described by a coupling
scheme in which {\bf I} and {\bf S} are coupled to form {\bf K}, and
then {\bf L} is coupled to {\bf K} to form {\bf F}.  In this limit,
the eigenvalue $K$ is approximately preserved as a good quantum
number.  The small deviations from the \textit{IS}-coupling scheme are well
accounted for by an exact diagonalization for the intermediate coupling case.
However, with increasing $n$, the \textit{IS}-coupling scheme should rapidly become
more accurate because the fine-structure interactions decrease in proportion
to $1/n^3$, while the hyperfine interactions tend to a constant at the
series limit.  The surprise is that it already works so well for
$n= 3$.

We would like to thank K. Bailey and T. P. O'Connor for technical support. 
This work was supported by the U.S. Department of Energy, Office of Nuclear Physics and Office of Basic
Energy Sciences, Office of Science, under Contract No.\ DE-AC02-06CH11357.  G.W.F.D. acknowledges support by the Natural
Sciences and Engineering Research Council of Canada, and by SHARCNET. 

\bibliography{He3Sulai}

\end{document}